\documentstyle[12pt,epsf]{article}
\def\half{\frac{1}{2}}
\def\be{\begin{equation}}
\def\ba{\begin{eqnarray}}
\def\ea{\end{eqnarray}}   
\def\gr{\nabla}
\def\ee{\end{equation}}
\def\to{\rightarrow}
\def\tmu{$TH\epsilon\mu\;$}
\def\pd{\partial}
\def\nn{\not\!}
\def\sp{\overline\psi}
\def\ov{\overline}
\def\api{\frac{\alpha}{\pi}}
\def\al{\alpha}

\def\om{\Omega}
\renewcommand{\thesection}{\Roman{section}}

\begin{document}
\baselineskip=0.6cm
\begin{titlepage}
\begin{center}
{\Large\bf  The Equivalence Principle
and Anomalous Magnetic Moment Experiments}\vspace{1cm}\\
C. Alvarez
\footnotemark\footnotetext{E-mail:
calvarez@avatar.uwaterloo.ca} and 
R.B. Mann\footnotemark\footnotetext{E-mail:
mann@avatar.uwaterloo.ca}\\
Department of Physics \\
University of Waterloo\\
Waterloo, ONT N2L 3G1, Canada\\
\vspace{2cm}
\today\\
\end{center}
\vspace{2cm}
\begin{abstract}
We investigate the possibility of 
testing of the Einstein Equivalence Principle (EEP)
using measurements 
of anomalous magnetic moments of elementary particles.
We compute the one loop correction
for the $g-2$ anomaly within the class of 
non metric theories of gravity described
by the \tmu formalism. 
We find several novel mechanisms for breaking the EEP whose origin
is due purely to radiative corrections. 
We discuss the possibilities of setting
new empirical constraints on these effects.

\end{abstract}
\end{titlepage}

\section{Introduction and Summary}

Metric theories of gravity offer the singular beauty of endowing 
spacetime with a symmetric, second-rank tensor field
 $g_{\mu\nu}$ that couples universally to all non-gravitational fields. 
This unique operational geometry is dependent upon 
the validity of the Einstein Equivalence Principle (EEP), which 
states that the outcomes of nongravitational test experiments
performed within a local, freely falling frame are independent 
of the frame's location 
(local position invariance, LPI) and velocity (local Lorentz 
invariance, LLI) in a background gravitational field.
Non metric theories break this universality by coupling additional 
gravitational fields to matter, and so violate either LPI, LLI or both.
Limits on LPI or LLI are imposed by gravitational redshift and atomic 
physics experiments respectively, by comparing atomic energy transitions 
that are sensitive to these symmetries. Laser experiments have set 
stringent limits on violations of LLI (to a precision of $\sim 10^{-22}$) 
\cite{PLC}, while the next generation of
gravitational redshift experiments could reach a 
precision up to $10^{-9}$ \cite{will1}.

These experiments probe transitions that are predominantly sensitive to
nuclear electrostatic energy, although violations of WEP/EEP due to
other forms of energy (virtually all of which are associated with baryonic
matter) have also been estimated \cite{HW}.  However there exist many 
other physical systems, dominated by primarily non-baryonic
energies, for which the validity of the EEP is comparatively less  
well understood \cite{Hughes}.  Included in the list of
such systems are photons of differing
polarization \cite{Jody}, antimatter systems \cite{anti}, 
neutrinos \cite{utpal}, mesons \cite{Kenyon}, massive leptons, 
hypothesized dark matter, second and third generation matter, and
quantum vacuum energies. 

In order to establish the universal behavior of gravity,
it is important to empirically confront the EEP with as diverse a 
range of non-gravitational interactions as is possible.
Amongst the most intriguing of the list of physical systems noted
above are those for which potential violations of the EEP can arise
in quantities dependent upon 
vacuum energy shifts, which are peculiarly quantum-mechanical in origin
({\it i.e.} do not have a classical or semi-classical
description). The empirical validity of the EEP
in physical systems where radiative corrections are non-negligible
remains an open question. Such systems provide an interesting empirical 
regime for a confrontation between gravitation and quantum mechanics.

Perhaps the most useful framework for
further such testing of the EEP is provided by
quantum electrodynamics (QED). We considered this approach
in a previous paper \cite{catlamb} 
by analyzing the behavior of Lamb shift transition energies 
within the context of a wide class of
nonmetric theories of gravity. The Lamb shift,  
along with the anomalous magnetic
moments ($g-2$ factors) of elementary particles, 
presents the most compelling evidence in support of QED. 
It is therefore 
natural to extend the nonmetric analysis to the $g-2$ anomaly.

We consider in this paper the possibility of using measurements 
of anomalous magnetic moments of elementary particles as a 
possible test of the EEP. 
The high precision attained in $g-2$ experiments motivated 
the early work of Newman {\it et al.} \cite{NFRS}
to use such experiments to set new bounds on the validity of 
special relativity. Similarly, we expect (and shall subsequently
demonstrate) that such experiments could impose stringent and qualitatively 
new limits on the parameter space of nonmetric theories.
Our interpretation of those experiments is substantially different from 
theirs as we 
explicitly include violations of the EEP in the computation of $g-2$.
These effects  were not considered in ref. \cite{NFRS}, 
which assumed that violations of special 
relativity arose only from the dynamical equation governing the motion
of the fermion.

We consider the class of non-metric theories described by
the \tmu formalism \cite{tmu}, following the approach given in 
ref. \cite{catlamb} in developing  gravitationally modified (GM)
QED. This formalism  
encompasses a wide class of nonmetric theories of gravity, and deals 
with the dynamics of charged particles and electromagnetic fields in a static, 
spherically symmetric gravitational field. It assumes that the 
(classical) non-gravitational laws of physics can be derived from an action:
\begin{eqnarray}\label{1}
 S_{NG}&=&-\sum_a m_a\int dt\, (T-Hv_a^2)^{1/2}+\sum e_a \int dt\, v_a^\mu
 A_\mu(x_a^\nu)\nonumber\\
& &+ \half\int d^4x\,(\epsilon E^2- B^2/\mu),
\end{eqnarray}
where $m_a$, $e_a$, and $x_a^\mu(t)$ are the rest mass, charge, and
world line of particle $a$, $x^0\equiv t$, $v_a^\mu\equiv dx_a^\mu/dt$,
$\vec E\equiv-\vec\gr A_0-\pd\vec A/\pd t$,
$\vec B \equiv \vec\gr\times\vec A$. The metric is assumed to be
\be\label{tmumet}
ds^2 = T(r)dt^2 - H(r)(dr^2+r^2d\Omega^2) 
\ee
where $T$, $H$, $\epsilon$, and $\mu$ are
arbitrary functions of the (background) Newtonian gravitational potential
$U= GM/r$, which approaches unity as $U\to 0$.  For an arbitrary non-metric
theory, these functions will depend upon the type of matter, {\it i.e.}
the species of particle or field coupling to gravity. The functions
$\epsilon$ and $\mu$ parameterize the `photon metric', whereas
$T$ and $H$ parameterize the `particle metric' in the static, spherically
symmetric case.  Although we shall generically employ the notation
$T$ and $H$ throughout this paper, it should be kept in mind that 
these functions shall in general have one set of values 
for electrons, another set for muons, another for protons, {\it etc.}.  
Universality of gravitational
coupling in the particle sector implies that the $T$ and $H$ functions 
are species independent.  It is an empirical question as to whether or
not such universality holds for all particle species.
The stringent limits on universality violation set by 
previous experiments \cite{PLC} have
only been with regards to the relative gravitational couplings in
the baryon/photon sector of the standard model. For 
the leptonic sector relevant to our considerations, relatively little
is known \cite{Hughes}.

In order to study atomic systems, this classical action can be
generalized for quantum mechanical systems \cite{gabriel},
\be\label{q1}
S=\int d^4x\, \sp(i\nn\pd+e\nn\! A-m)\psi
 + \half\int d^4x\,(E^2-c^2B^2),
\ee
where local natural units are used, $\nn\! A=\gamma_\mu
A^\mu$, and $c^2=H_0/T_0\epsilon_0\mu_0$ with
the subscript ``0'' denoting the functions evaluated at $\vec X=0$,
the origin of the local frame of reference. 
This action emerges upon replacing the point-particle part of the
action in (\ref{1}) with the Dirac Lagrangian, expanding the \tmu
parameters about the origin,  neglecting their spatial variation over 
atomic distance scales, and rescaling coordinates and fields. 
These operations have the effect of treating spatial and temporal
derivatives and gauge couplings in the fermionic sector on the same
footing, and so we need not add additional terms of the form
$\sp(i\gamma^0(\pd_0+eA_0))\psi$ times an arbitrary constant, as such
terms can be re-absorbed into a definition of the parameters which appear
in (\ref{q1}) above, as noted below.

The action (\ref{q1}) refers to the preferred frame, as defined by the
rest frame of the external gravitational field $U$. In order to
analyze effects in systems moving with velocity $\vec u$
with respect to that frame
we need to transform the fields and coordinates in (\ref{q1}) via the
corresponding Lorentz transformations. This gives
\begin{eqnarray}\label{q2}
S&=&\int d^4x\, \sp(i\nn\pd+e\nn\! A-m)\psi + \int d^4x\, J_\mu A^\mu
\nonumber\\
 &+&\half\int d^4x\,\left[(E^2-B^2)\right.\\
 &+& \left.\xi\gamma^2\left(\vec u^2 E^2-(\vec u\cdot\vec E)^2+B^2-(\vec u\cdot\vec B)^2
  +2\vec u\cdot(\vec E\times\vec B)\right)\right]\nonumber.
\end{eqnarray}
where  $J^\mu$ is the electromagnetic 4-current
associated with some external source, $\gamma^2=(1-\vec u^2)^{-1}$, 
and $\xi=1-c^2$ is a dimensionless (and species-dependent) parameter 
that measures the degree to which LPI/LLI is broken. This parameter 
scales with the magnitude of the dimensionless Newtonian potential,
which is expected to be much smaller than unity for actual experiments.
We are therefore able
to compute effects of the terms in eq. (\ref{q2}) that break local 
Lorentz invariance via a perturbative analysis about the familiar and
well-behaved $c\to 1$ or $\xi\to 0$ limit. 

We therefore consider gravitationally modified
Quantum Electrodynamics (GMQED) based on the  action (\ref{q2}).
The fermion sector and the interaction term do not change with respect
to the metric case, and so neither do the fermion propagator and the
vertex rule. All the non-metric effects are accounted for
in the pure electromagnetic sector of the action, which in turns modifies
the photon propagator. After a proper choice of the gauge fixing term
involved in the quantization of the photon field, it can be shown \cite{catlamb} that the 
photon propagator takes the form up to first order in $\xi$ (in momentum space): 
\begin{equation}\label{11} 
G_{\mu\nu}=-(1+\xi)\frac{\eta_{\mu\nu}}{k^2}
+\xi\frac{\gamma^2}{k^2}\left[\eta_{\mu\nu}\frac{(\beta\cdot k)^2}{k^2}
+\beta_\mu\beta_\nu\right] ,
\end{equation}
where $\eta_{\mu\nu}$ is the Minkowski tensor with a signature (+ - - -);
$\gamma^2\equiv 1/(1-\vec u^2)$
and  $\beta^\mu\equiv(1,\vec u)$;
henceforth $\beta^2\equiv 1-\vec u^2$.

Therefore eq. (\ref{11}), along with the unmodified fermion 
propagator $S_F(p)$ and vertex rule, may be used as the basis of the
Feynman rules of GMQED. 
The radiative corrections affecting these quantities  are defined
in terms of the photon self energy $\Pi^{\mu\nu}(k)$, 
fermion self energy $\Sigma(p)$, and
vertex function $\Gamma^\mu$ respectively. These insertions involve 
the calculation of loop integrals as given by the 
Feynman rules up to a given order. 

The addition of more parameters to the theory also entails new
renormalizations beyond those of the wavefunctions,  
charge and mass of the fermion. The \tmu parameters appear as functions
of  $c_0^2\equiv T_0/H_0$ and $c_*^2\equiv 1/\mu_0\epsilon_0$,
and must then  be correspondly redefined. 
In units where $c_0\equiv 1$ ($c_*=c$), EEP-violating corrections only 
appear in the electromagnetic sector of 
the action (as terms proportional to $\xi$). However a more general choice
is $c_0\not =1$, for which the particle sector of the Lagrangian 
density is of the form 
\be
{\cal L}_D=\sp(\nn p-\nn V-m)\psi+\xi_0\sp(p_0-V_0)\gamma^0\psi
\ee
with $\xi_0\equiv 1-c_0^{-1}$. In the moving frame this is 
(up  to a constant)
\ba \label{apbD}
{\cal L}'_D&=&\sp(\nn p-\nn V-m)\psi\\
&+&\xi_0\gamma^2\sp(\beta\cdot p-\beta\cdot V)\nn \beta\psi \quad .
\nonumber
\ea
{}From (\ref{apbD}) we see that quantum corrections of the form
\be
\delta {\cal L}_D=\sp(\delta\xi_0^{(1)}\beta\cdot p-\delta\xi_0^{(2)}
\beta\cdot V)\nn\beta\psi
\ee
can still be expected. It is straightforward to show 
that gauge invariance will guarantee $\delta\xi_0^{(1)}=
\delta\xi_0^{(2)}=\delta\xi_0$. 

Hence, in order to renormalize the mass and the \tmu parameters, we have to
include counterterms of the form
\be\label{coun1}
\delta m+\delta\xi_0\nn\beta(\beta\cdot p-\beta\cdot V),
\ee
which consequently participate in the redefinition of the fermion 
self energy and vertex function.
Additionally, the electromagnetic sector will induce quantum fluctuations
of the form:
\be\label{coun2}
\delta {\cal L}_{EM}=\delta \xi A^\mu\{(k^2-\gamma^2(\beta\cdot k)^2)\eta_{\mu\nu}
-\gamma^2\beta_\mu\beta_\nu k^2\}A^\nu
\ee
corresponding to the renormalization of the \tmu parameters,  
or equivalently $\xi\equiv 1-H_0/T_0\mu_0\epsilon_0$, entailing 
a renormalization of the photon self energy.

This summarizes the procedure for performing perturbative calculations in 
GMQED as employed previously in ref. \cite{catlamb} in computing
the Lamb shift. This calculation was complicated by
the boundedness of the electron to the hydrogenic atom under consideration.
Here, in the $g-2$ case, we have to deal only with free leptonic states. 
In Sec. II we evaluate the (one loop) radiative corrections 
to the elastic scattering of a free electron by an external 
electromagnetic field. 
Sec. III relates the scattering amplitude with experimental observables 
describable in terms of the g-2 anomaly. This requires a
derivation of the relativistic equation of motion for the electron spin 
in the presence
of a magnetic field. This follows from a classical treatment for the electron
and magnetic field. Quantum effects are introduced in the modified
Hamiltonian only (as are non metric effects). The connection with possible
LLI/LPI violating experiments is presented at the end of this section.
Discussion of the results and general comments are given in Sec. IV.
More details about the loop calculation are shown in the Appendix.  

\section{(GM) Free Scattering}

We shall consider the lowest order radiative correction to
the elastic scattering of 
electrons by a static external field $A^\mu$. These one loop
contributions can be summarized in terms of the Feynman diagrams
illustrated in Fig. 1.
\begin{figure}[ht]
\centering
\leavevmode
\epsfbox[90 80 280 360] {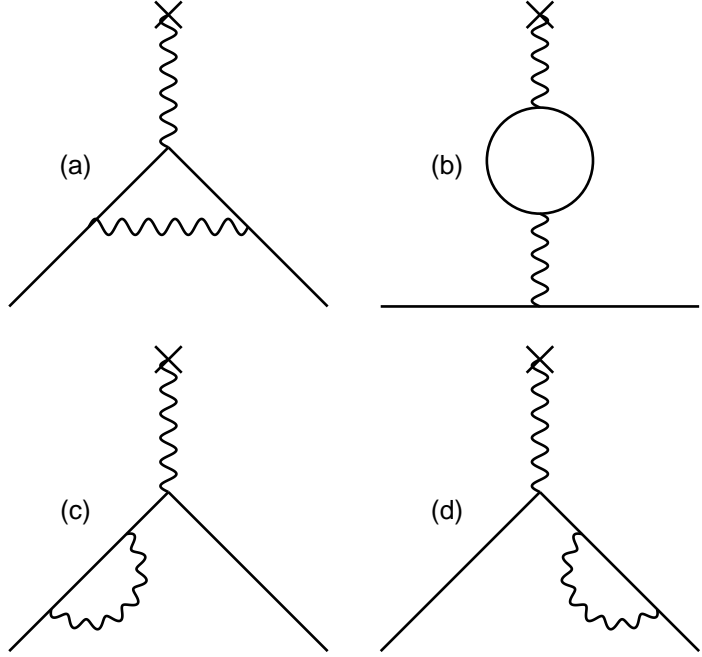}
\caption{One loop corrections to the elastic scattering 
of an electron by an external electromagnetic source}
\end{figure}

The Feynman amplitudes for the diagrams follow from the Feynman rules giving the result \cite{M&S}:
\be\label{ampli}
\Lambda ^\mu(p',p)=\ov u(\vec p\,')\left\{\Gamma^\mu+P^\mu+L^\mu \right\}u(\vec p)
\ee
where
\ba 
\label{vertex}
\Gamma^\mu(p',p)&=&\frac{(ie)^2}{(2\pi)^4}\int d^4k\gamma^\al 
iS_F(p'-k)\gamma^\mu iS_F(p-k)
\gamma^\beta i G_{\al\beta}(k)\\
\label{pola}
P^\mu(p',p)&=&\gamma^\al i G_{\al\beta}(q)i \Pi^{\beta\mu}(q)\\
\label{leg}
L^\mu(p',p)&=&i\Sigma(p')iS_F(p')\gamma^\mu+\gamma^\mu iS_F(p)i\Sigma(p)
\ea
with
\ba\label{self}
i\Sigma(p)&=&\frac{(ie)^2}{(2\pi)^4}\int d^4k i G_{\al\beta}(k)\gamma^\al iS_F(p-k)\gamma^\beta\\
\label{pola1}
i\Pi^{\beta\mu}(q)&=&\frac{(ie)^2}{(2\pi)^4}(-)\hbox{Tr}\int d^4k\gamma^\beta iS_F(k+q)
\gamma^\mu iS_F(k)
\ea
and $q\equiv p'-p$.

We refer to eqs. (\ref{vertex}),  (\ref{pola}), and (\ref{leg})
as the Vertex, Polarization, and Leg
contributions, which respectively correspond to diagrams
(a), (b) and (c) plus (d). We also note that  
expressions (\ref{vertex}), (\ref{self}),
(\ref{pola1}) represent the one loop corrections to the vertex, fermion and
photon self energy parts respectively.

Given the form of the photon propagator it is convenient to introduce:
\be
\Lambda^\mu=(1+\xi)\Lambda^\mu_0+\gamma^2\xi\Lambda^\mu_\xi
\ee
where the subscript ``0" denotes the (known) result coming from the standard
part of the photon propagator, 
and ``$\xi$" for the part proportional to $\gamma^2$ in (\ref{11})
\be\label{fp}
G_{\mu\nu}^\xi=\frac{\beta_\mu\beta_\nu}{k^2}
+\eta_{\mu\nu}\frac{(\beta\cdot k)^2}{k^4}
\ee
In the remainder of this section 
we consider this part of the propagator only,
omitting the ``$\xi$" label in the corresponding expressions.

The procedure for evaluating the loop integrals is equivalent 
to that of standard (or metric)
QED. We need to regularize them first and then renormalize the 
parameters,
which include the \tmu parameter along with the fermion charge and mass.
The regularization of the photon propagator is carried out using
\be\label{regu}
\frac{1}{k^2}\to-\int_{\mu^2}^{\Lambda^2}\frac{dL}{(k^2-L)^2},\qquad
\frac{1}{k^4}\to-2\int_{\mu^2}^{\Lambda^2}\frac{dL}{(k^2-L)^3},
\ee
with the assumed limits $\mu\to 0$ and $\Lambda\to \infty$,
and the parameter renormalization by the inclusion of the corresponding 
counterterms
to each loop integral. Details about this procedure and the 
corresponding calculations are given
 in the appendix. We quote the final result for the loop integrals:
\ba\label{self-f}
\Sigma(p)&=&\api(\nn p-m)\left\{\frac{(\beta\cdot p)^2}{m^2}
\frac{2}{3}-\beta^2\left[\frac{5}{24}\ln(\frac{\Lambda}{m})^2+\frac{133}{144}-\frac{1}{4}
\ln(\frac{m}{\mu})^2\right]
\right\}\\\nonumber 
&-&\api\frac{\beta\cdot p}{6m}\left\{(\nn p-m)\nn\beta+\nn\beta(\nn p-m)\right\}
+O\Big((\nn p-m)^2\Big)\\\label{vertex-f} 
\Gamma^\mu(p',p)&=&\api\Big\{\gamma^\mu\left[
\frac{2}{3}\frac{(\beta\cdot p)^2}{m^2}
-\beta^2\left(\frac{5}{24}\ln(\frac{\Lambda}{m})^2+\frac{133}{144}-\frac{1}{4}
\ln(\frac{m}{\mu})^2\right)\right.\nonumber\\
&+&\left[\frac{q^2}{m^2}\left(\frac{17}{144}\beta^2+\frac{1}{12}+
\ln(\frac{m}{\mu})^2(\frac{1}{8}-\frac{\beta^2 }{24})\right)\right.\nonumber\\
&+&\left.\frac{3}{2}\frac{\beta\cdot p}{m}\frac{\beta\cdot q}{m}
+(\frac{\beta\cdot q}{m})^2\left(\frac{19}{36}-\frac{1}{6}\ln(\frac{m}{\mu})^2\right)\right]\nonumber\\
&-&\frac{\nn q}{m}\gamma^\mu\nn\beta\left(\frac{1}{6}\frac{\beta\cdot p}{m}
+\frac{1}{12}\frac{\beta\cdot q}{m}
\right)+\left(\frac{1}{9}\frac{q^2}{m^2}-\frac{1}{3}\right)\nn\beta\beta^\mu
\\\nonumber&-&\frac{1}{6}\frac{\nn q}{m}\nn\beta\beta^\mu
-\half\frac{\beta\cdot q}{m}\nn\beta\gamma^\mu+\frac{2}{3}\frac{\beta\cdot q}{m}\beta^\mu+
\left(\frac{\beta^2}{24}+\frac{1}{6}+\frac{1}{6}\frac{\beta\cdot p}{m}\frac{\beta\cdot q}{m}\right)
\frac{\nn q}{m}\gamma^\mu
\\\nonumber&+&\left(\frac{1}{6}\frac{\beta\cdot p}{m}-\frac{5}{36}\frac{\beta\cdot q}{m}
\right)\frac{q^\mu}{m}\nn\beta
-\left(\frac{\beta^2}{24}+\frac{1}{6}\right)\frac{q^\mu}{m}\Big\}+O(q^3)\nonumber\\\label{pola-f}
\Pi^{\al\beta}(q)&=&-\api\left(q^2\eta^{\al\beta}-q^\al q^\beta\right)\frac{q^2}{15m^2}+O(q^6)
\ea  
where we have implicitly assumed that  (\ref{vertex-f}) is acting on free spinors.

The Ward identity 
\be\label{Wardi}
\frac{\pd\Sigma(p)}{\pd p_\mu}=\Gamma^\mu(p,p).
\ee
is a consequence of gauge invariance, 
and therefore it holds even in the absence of Lorentz invariance. It is
straightforward to check that (\ref{self-f}) and (\ref{vertex-f}) 
satisfy (\ref{Wardi}).

The evaluation of (\ref{ampli}) is also
straightforward once the loop integrals have been calculated.
We just comment on the computation of the Leg correction, 
which is ambiguous since it contains terms 
like $``0/0"$, which are indeterminate. To obtain
an unambiguous result, we must explicitly introduce a 
damping factor, which is necessary for the correct
definition of the initial and final ``bare'' states. 
Details of this adiabatic approach are presented in appendix B.
The final result for the Leg correction is
\ba
L^\mu&=&\api\Big\{\gamma^\mu\left[\frac{2}{3}\left(\frac{(\beta\cdot p)^2}{m^2}+
\frac{\beta\cdot p}{m}\frac{\beta\cdot q}{m}
+\half\frac{(\beta\cdot q)^2}{m^2}\right)\right.\nonumber\\
&+&\left.\beta^2\left(\frac{5}{24}\ln(\frac{\Lambda}{m})^2+\frac{133}{144}-\frac{1}{4}
\ln(\frac{m}{\mu})^2\right)\right]\\\nonumber
&+&\frac{1}{3}\nn\beta\beta^\mu-\frac{1}{6}\frac{\nn q}{m}\nn\beta\beta^\mu+\frac{1}{6}
\frac{\beta\cdot q}{m}\nn\beta\gamma^\mu\Big\} 
\ea

Note that this part gives a contribution to the total amplitude that 
{\it cannot} be removed after
renormalization. Furthermore, the gauge invariance of the Feynman 
amplitude which  is manifest as
\be
q\cdot \Lambda=0
\ee
requires the presence of such terms,
a condition that is not satisfied by the vertex contribution only. 

 The final result for the scattering amplitude is
\be\label{final}
\Lambda^\mu=F^\mu+G^\mu+I^\mu
\ee
with
\ba\label{ff}
F^\mu&=&\api\Big\{\gamma^\mu\left[
\frac{q^2}{m^2}\left(\frac{17}{144}\beta^2+\frac{1}{12}+
\ln(\frac{m}{\mu})^2(\frac{1}{8}-\frac{\beta^2 }{24})\right)\right.
+\frac{5}{6}\frac{\beta\cdot p}{m}\frac{\beta\cdot q}{m}
\nonumber\\&+&\left.(\frac{\beta\cdot q}{m})^2\left(\frac{47}{180}-\frac{1}{6}\ln(\frac{m}{\mu})^2\right)\right]
-\frac{\nn q}{m}\gamma^\mu\nn\beta\frac{1}{12}\frac{\beta\cdot q}{m}
\\&+&\nonumber\frac{8}{45}\frac{q^2}{m^2}\nn\beta\beta^\mu
+\frac{1}{6}\frac{\beta\cdot p}{m}\frac{\beta\cdot q}{m}\frac{\nn q}{m}\gamma^\mu\Big\}
\\\label{gg}
G^\mu&=&\api\Big\{-\frac{\nn q}{m}\gamma^\mu\nn\beta\frac{1}{6}\frac{\beta\cdot p}{m}
-\frac{1}{3}\frac{\nn q}{m}\nn\beta\beta^\mu-\frac{1}{3}\frac{\beta\cdot q}{m}\nn\beta\gamma^\mu
+(\frac{\beta^2}{24}+\frac{1}{6})\frac{\nn q}{m}\gamma^\mu\Big\}\\\label{ii}
I^\mu&=&\api\Big\{\frac{2}{3}\frac{\beta\cdot q}{m}\beta^\mu
+\left(\frac{1}{6}\frac{\beta\cdot p}{m}-\frac{37}{180}\frac{\beta\cdot q}{m}
\right)\frac{q^\mu}{m}\nn\beta
-\left(\frac{\beta^2}{24}+\frac{1}{6}\right)\frac{q^\mu}{m}\Big\}
\ea

The various terms in (\ref{final})  distinguish the different 
contributions to the scattering amplitude. In (\ref{ff}) we group 
terms of order $q^2$ or higher. $G^\mu$
accounts for terms of order $q$ at least, and $I^\mu$ for the gauge terms or 
those who  give no contribution to the amplitude. 
Note that the remaining infrared divergence
in $F^\mu$ can be understood in terms of soft photon 
radiation, analogous to  the metric case.

In the next section we will use the above results to compute
the $g-2$ anomaly.

\section{(GM) g-2}

To lowest order the Feynman amplitude associated with
the elastic scattering of an electron by a static external field
is 
\be\label{zero}
ie\ov u(p')\nn A(q)u(p) \quad.
\ee
The radiative correction of order $\al$ to this process is given by
\be\label{uno}
ie\ov u(p')\{(1+\xi)\Lambda_0\cdot A+\gamma^2\xi\Lambda_\xi\cdot A\}u(p)
\ee
where $\Lambda_0$ represents the (known) metric 
result and $\Lambda_\xi$ represents the contribution from
(\ref{final}).

In the nonrelativistic limit of slowly moving particles ($|\vec q|\to 0$) and a static
magnetic field , it is straightforward to show that
\ba
e\nn A(q)&\to& -\frac{e}{2m}\vec B\cdot\vec\sigma\\
e\Lambda_0\cdot A&\to&-\frac{e}{2m}(\frac{\al}{2\pi})\vec B\cdot\vec\sigma\\
e\Lambda_\xi\cdot A&\to&e\,G\cdot A \label{xigg}
\ea
with $G^\mu$ given by (\ref{gg}), which is the dominant term as $q\to 0$.

In order to simplify this contribution, 
we consider a constant  magnetic field
$\vec B$, that is $\vec A=\half\vec r\times\vec B$, in which case
\be
\nn q\nn\beta\beta\cdot A\to-\half(\vec B\cdot\vec u\,\vec\sigma\cdot\vec u
-\vec B\cdot\vec\sigma\,\vec u^2)
\ee
where we have neglected the terms that mix the 
large and small spinor components. Similarly, we can show
\be
\beta\cdot q\nn\beta\nn A\to-\half(\vec B\cdot\vec u\,\vec\sigma\cdot\vec u
-\vec B\cdot\vec\sigma\,\vec u^2)
\ee
and in the non relativistic limit
\be
\nn q\nn A\nn\beta\frac{\beta\cdot p}{m}\simeq\nn q\nn A\nn\beta
\to-\vec B\cdot\vec\sigma
\ee
and
\be
\nn q\nn A\to-\vec B\cdot\vec\sigma \quad .
\ee

If we put everything together in (\ref{gg}):
\be
e\,G\cdot A\to-\frac{e}{2m}\frac{\al}{\pi}\{\vec B\cdot\vec\sigma(\frac{1}{12}+\frac{7}{12}\vec u^2)
-\frac{2}{3}\vec B\cdot\vec u\,\vec\sigma\cdot\vec u\Big\}
\ee
As a cross-check on the above result, we take 
the limit $u_iu_j\to-\delta_{ij}$ obtaining
$G\cdot A\to-2\Lambda_0\cdot A$, which is the required limit
consistent with the structure of  eq. (\ref{fp}) in that case.
The previous  result is the  contribution of (\ref{xigg}) to (\ref{uno}),
which added to (\ref{zero}), give us  the
relevant part of the Hamiltonian as
\be\label{inter}
 H_\sigma=-\{\Gamma\,\vec S\cdot\vec B+\Gamma_*\vec S\cdot\vec u\,\vec B\cdot\vec u\}
+O(\xi^2)O(\al^2)
\equiv -\Gamma^{ij}S_iB_j
\ee
with
\ba\label{gs}
\Gamma\equiv\frac{e}{2m}g&\equiv&\frac{e}{2m}\{2+
\api[1+\xi(1+\frac{\gamma^2}{6}(1+7\vec u^2))]\}
\\\label{gu}
\Gamma_*\equiv\frac{e}{2m}g_*&\equiv&-\frac{e}{2m}\api\xi\frac{4}{3}\gamma^2
\ea
where we have identified $\vec S\equiv\frac{\vec\sigma}{2}$, and 
$\hat u=\vec u/|\vec u|$. The $\Gamma$
parameters account for the coupling strength 
between the magnetic field and spin. We see that $\Gamma_{ij}$ generalizes
the gyromagnetic ratio of a fermion analogous to the manner in which the
anomalous mass tensor generalizes the mass of a particle \cite{haugan}.
We therefore identify the parameters $\Gamma^{ij} \equiv
\Gamma \delta^{ij} + \Gamma_* u^i u^j$ 
with the components of the {\it anomalous gyromagnetic ratio tensor} 
of the fermion in the class of \tmu theories.

Note that the presence of preferred frame
effects induces a qualitatively new form of interaction 
between the spin and magnetic field which is
quantified by $\Gamma_*$. Here, instead of coupling with each other, they
both couple independently to the fermion
velocity relative to the preferred frame. This interaction 
stems purely from radiative corrections, and would be absent in any
tree-level analysis of GMQED.

Hence, eq. (\ref{inter}) describes the interaction (as seen from
the particle rest frame) between the particle spin
and an external homogeneous magnetic field. From this we can extract 
the energy difference between electrons with opposite
spin projection in the direction of the magnetic field as:
\be\label{eshift}
\Delta E_\sigma=-\frac{eB}{2m}\Big [ g+g_*u^2\cos^2\Theta\Big] 
\ee
where $\Theta$ is the angle between the magnetic field and the preferred frame velocity.
The influence of the radiative corrections (coming from $g-2$ and $g_*$) 
in (\ref{eshift}) is negligible
in comparison to the dominant factor of 2 in $g$.
Since we want to single out the effects of the non-metric
corrections, it is more interesting to study the
precession of the spin or, more specifically, the oscillation of the longitudinal
spin polarization. In the metric case, this frequency is 
proportional to the factor $g-2$, and so
it is a distinctive signature of radiative corrections.

The observable quantity in the $g-2$ experiments is actually the electron
polarization, which is proportional to the quantum mechanical 
expectation value of $\vec S$, that is, $\langle \vec S\rangle$. 
Using Ehrenfest's theorem, a quantum
mechanical solution for the motion of $\langle \vec S\rangle$ is obtained  
from the equation
\be\label{spin1}
\frac{d\vec S}{dt'}|_{R.F.}=-i[\vec S, H_\sigma]=\vec S\times\Big[\Gamma\vec B'+
\Gamma_*(\vec B'\cdot\vec u)\vec u\Big]
\ee 
where the primed variables are referred explicitly to the particle rest 
frame ($R.F.$). Note that the preferred frame effect will show distinctly 
as a temporal variation of the spin component parallel to the magnetic 
field. 

In general we want to know the spin precession relative 
to some specific laboratory system, with respect to which
the particle is moving with some velocity $\vec\beta$.
This frame need not {\it a-priori}  be the previously defined
preferred frame, and so $\vec\beta\not =\vec u$.

Since the \tmu formalism does not change (locally) the 
fermion electromagnetic
field interaction, we  expect that a charged particle 
in the presence of an homogeneous magnetic field
will satisfy the equation
\be\label{spin2}
\frac{d\vec\beta}{dt}=\vec\beta\times\vec \om_c
\ee
with the cyclotron frequency $\vec \om_c=\frac{e}{m\gamma}\vec B$ 
and $\gamma=(1-\vec\beta^2)^{-1/2}$.
Relating (\ref{spin1}) to the laboratory system  yields
\be\label{spin3}
\frac{d\vec S}{dt}|_{Lab}=\frac{d\vec S}{dt}|_{R.F.}
+\vec \om_T\times\vec S
\ee
due to Thomas precession, with
$\vec \om_T=\frac{\gamma^2}{\gamma+1}(\frac{d\vec\beta}{dt}\times\vec\beta)$.
This frequency is  kinematic in origin 
and it is a consequence of the non-commutativity
of the Lorentz transformations.

Relating the primed variables in (\ref{spin1}) to
the laboratory ones by a Lorentz transformation gives
\be\label{spin5}
\frac{d\vec S}{dt}|_{Lab}=\vec S\times\vec\om_s
\ee
with
\be
\vec\om_s=\Gamma\vec B+(1-\gamma)\vec\om_c+
\Gamma_*(\vec B\cdot\vec u)\vec u
\ee
where we have set $\vec E=0$ and considered (for simplicity) the 
case of orbital motion perpendicular
to the magnetic field  ($\vec\beta\cdot\vec B=0$) in the
above. Note that the spin precession about $\vec \om_s$ is no longer
parallel to the magnetic field (axial direction), but has a component
parallel to $\vec u$ that comes from radiative and nonmetric effects.

At this point it becomes necessary to define the preferred coordinate system.
There are several candidates (such as the rest frame of the cosmic 
microwave background) for this frame \cite{will1}. 
To study this issue it is sufficient
to assume that  the laboratory system (Earth) moves with a
non-relativistic velocity ($\vec V$) with respect to the preferred
frame, and so we can identify 
$$\vec u=\vec V+\vec\beta \qquad .$$

In order to single out the effects of radiative corrections, we study
the spin precession relative to the rotational motion of the
electron, that is:
\be\label{set1}
\frac{d\vec S}{dt}|_{rot}=\vec S\times\vec\om_D
\ee
with $\vec\om_D=\vec\om_s-\vec\om_c$  and  $\vec S=(S_\bot^{\|},S_\bot^{\bot},S_\|)$,
 where the first two components are perpendicular to $\vec B$ (lower index) but
parallel and perpendicular to $\vec\beta$ (upper index), and the last one parallel to $\vec B$.
In the following we refer to the difference
frequency  ($\om_D$)  as the anomalous frequency (given its connection with 
the anomalous magnetic moment in the metric case). 
It is convenient to rewrite:
\be
 \vec\om_D=\vec \om_a+\om_a^*\cos\Theta (\vec V_\bot+\vec\beta)
\ee
with
\be\label{anomal}
\vec\om_a=\frac{e}{2m}\left(g+g_*V^2\cos^2\Theta-2\right)\vec B
\ee
and $\om_a^*=\frac{e}{2m}g_*BV$;
 where $\Theta$ represents the
angle between  $V$ and the magnetic field, and $V_\bot$ the component
of the velocity perpendicular to $B$. In $\om_a$ we group all the
terms parallel to the magnetic field that  contribute to the anomalous
frequency (including nonmetric effects). The remaining  
terms  perpendicular to $B$ arise from nonmetric effects only, and produce
a temporal variation of the 
spin component parallel to the magnetic field. This effect
is absent in the metric case, and so represents a qualitatively new
manifestion of possible EEP violation.

In general we are interested in solving (\ref{set1}) for the cases $\beta>>V$ or
$\beta<<V$ so that $\gamma(u)\simeq\gamma(\beta)$ or $\gamma(V)$, but
is otherwise constant.  Since
 $\om_a^*$ is proportional to $\xi$, we can  perturbatively solve for each component 
in (\ref{set1}). Taking, for example, 
the initial condition $\vec S(0)=S\hat\beta$ we find
\ba\label{set4}
S_\bot^\|&=&S\cos \om_at \qquad
S_\bot^\bot=S\sin \om_at\\\nonumber
S_\|&=&S\frac{\om_a^*}{\om_a}\beta\cos\Theta(1-\cos \om_at)+S\frac{\om_a^*}{\om_a+\om_c}
V\frac{\sin 2\Theta}{2}\left[\cos(\om_a+\om_c)t-1\right]
\ea
where we have chosen  a coordinate  system where $\hat B=\hat z$
so that 
\be
\hat V=\hat B\cos \Theta+ \hat x \sin\Theta,
\quad \hat\beta=\hat y \cos\om_ct-\hat x \sin\om_ct 
\ee
and assumed that any rotation  related to $\Theta$ is negligible
in comparison to  other frequencies involved in the 
problem ($\om_a$ or $\om_c$).
 
The fact that $\om_a$ was (in the metric case) proportional to
$g-2$, motived the very precise $g-2$ experiments which
were designed to specifically measure  that anomalous frequency. We see
that this frequency is modified by from its metric value by the additional
terms present in (\ref{anomal}). If we assume that the  EEP-violating 
contributions to  $\om_a$ are
bounded by the current level of precision for anomalous magnetic moments
\cite{Kino}, then the discrepancy between the best empirical 
and theoretical  values for the electron yields  the bounds
\be \label{belectron}
|\xi_{e^{-}}| < 3.5\times 10^{-8} \quad\mbox{and}\quad
|\xi_{e^{-}} - \xi_{e^+}| < 10^{-9}
\ee
the latter following from a
comparison of positron and electron magnetic moments.
For muons, a similar analysis yields
\be\label{bmuon}
|\xi_{\mu^{-}}| < 10^{-8}
\quad\mbox{and}\quad
|\xi_{\mu^{-}} - \xi_{\mu^+}| < 10^{-8}  \qquad   .
\ee
Even though the accuracy of the muon anomaly is lower than the
electron one, the slightly stronger bound in (\ref{bmuon}) arises
because the
experiments are carried out for high-velocity muons \cite {Bailey}.
To our knowledge these bounds on violation of gravitational
universality are the most stringent yet noted
for leptonic matter.

Newman {\it et. al.} analyzed the $g-2$ experiments
\cite{NFRS} in order to find
new bounds for the validity of special relativity.
They assumed that the parameter
$\gamma$ involved in the electron motion  had a different value
($\tilde\gamma$) from that which arises kinematically
(in Thomas precession and Lorentz transformations).
The equivalent equation for (\ref{anomal}) is in that case
\be
\Omega_a^{NFRS}=\frac{e B}{m}\left(\frac{g}{2}-\frac{\gamma}{\tilde\gamma}\right)
\ee
and by comparing with  two electron $g-2$ experiments,
 one at electron relativistic energy ($\beta=0.57$)
and the other nearly at rest ($\beta=5\times 10^{-5}$),
they obtained the constraint
$\delta \gamma/\tilde\gamma<5.3\times10^{-9} $ .
Our approach is qualitatively
different from theirs, in that we assume $\gamma=\tilde\gamma$
but include preferred frame effects in the evaluation of
the anomalous magnetic moment. A similar analysis in our case yields
the weaker bounds of
 $|\xi_e|<7\times 10^{-6}$ for electrons, and $|\xi_\mu|<2\times 10^{-7}$
for muons. In the latter we used the  $g-2$ muon experiments carried
at  $\beta=0.9994$ ($\gamma=29$)\cite{Bailey}, and
$\beta=0.92$ ($\gamma=12$)\cite{Bailey2}.

Preferred effects not only modify the anomalous frequency 
according to (\ref{anomal}), but also induce oscillations in 
the spin component parallel to $B$. As stated above, this is a qualitatively
new consequence of EEP violations due solely to radiative corrections
in GMQED. Searching for such oscillations therefore provides a
new null test of the EEP. We can estimate the magnitude of such effects 
by taking the temporal average
of $S_\|$ over the main oscillation given by $\om_a$, which gives
\be
\delta=\frac{\langle S_\|\rangle}{S}\sim\xi V\beta\cos\Theta\gamma^2
\ee
This effect is enhanced in highly relativistic situations, and can be
estimated by considering a
typical experiment with $V\sim 10^{-3}$. For 
electrons $\beta\sim 0.5$, and so $\delta_e\sim 10^{-11}$;
for muons $\beta=0.9994$, yielding $\delta_\mu\sim 10^{-8}$.
In both cases we used the corresponding present constraints for
$\xi$ given above.

The novelty of the $S_\|$ oscillation suggests the possibility
of putting tighter constraints on the non-metric parameter,
once appropriate experiments are carried out. The same goes for 
the analysis  of $\om_a$ at different values of $\Theta$ 
(the angle between the magnetic field and the velocity of
the laboratory system with respect to the preferred frame). The rotation
of the Earth will turn this orientation dependence into a time-dependence
of the anomalous magnetic moment, with a period related to that of the
sidereal day. 

The previous analysis was concerned with effects related to spatial
anisotropy.   We turn now to considering possible violations of 
local position invariance. The position dependence in the
former section was implicit in the redefinitions of charge, mass and fields. These quantities were
rescaled in terms of the \tmu functions, which were considered constant throughout the computation.
LPI violating experiments are of two types.  One of these entails
the measurement of a given frequency at two 
different points  in a gravitational field (where differences
in the gravitational potential could be significant)
within the same reference system. The other type involves
a comparison of frequencies arising from two different forms of
energy ({\it i.e.} two different clocks) at the same point in a
gravitational potential.
We parameterize the gravitational dependence on a given frequency
as:
\be\label{para1}
\om=\om^0\Big[1-U+\Xi^{ij}U_{ij}]+\cdots
\ee
where $U_{ij}$ represents the external gravitational tensor, satisfying 
$U_{ii}=U$, and the ellipsis represents higher order
terms (going as either $U^2$ or velocity times U) 
in the gravitational potential or terms independent
of it. 

The measured redshift parameter related to this frequency may 
be written as
\be
Z=\Delta U\Big(1-\Xi\Big),\qquad \Xi=\Xi^{ij}\frac{\Delta U_{ij}}{\Delta U}
\ee
where $\Xi^{ij}$ will depend upon the specific frequency measured in the experiments. Note
that this tensor is equivalent to  the anomalous passive gravitational mass tensor introduced for the
study of atomic transitions.

In $g-2$ experiments the relevant frequency is $\om_a$, which describes the precession of the longitudinal
polarization in the presence of a constant magnetic field. 
Using the \tmu formalism  (see eq.(\ref{anomal})) we obtain
\be\label{anor1}
\om_a=\frac{eB}{2m}\left[g-2\right]+\cdots=\frac{eB}{2m}\frac{\al}{\pi}\left[1+\xi\frac{7}{6}\right]+\cdots
\ee
where we have omitted terms proportional to  velocities, 
which eventually will contribute as $O(v^2U)$ terms at most.

In order to carry out the loop calculation, the \tmu dependence was
absorbed into the definition of the parameters under the rescaling
\be\label{rescal}
\om\to\om/c_0\qquad m\to m\sqrt{T_0}/c_0\qquad\al\to\al/\epsilon_0c_0
\ee
with $c_0=(T_0/H_0)^{1/2}$ as the limiting speed of the massive particles,
the subscript `0' denoting the \tmu functions evaluated locally at
$\vec{X}=0$. Although the product $eB$ remains invariant under this rescaling, 
the expression for the constant magnetic field still depends on 
the \tmu parameters once it is
written solely in terms of atomic parameters. This can be seen clearly by
considering the magnetic field produced by a long solenoid of length
$L$, with $N$ turns and carrying a current $I$. The gravitationally modified
Maxwell equation to solve is:
\be
\vec\gr\times (\mu^{-1}\vec B)=4\pi\vec J
\ee
and so we find the non-vanishing magnetic field inside the solenoid to be
$B=4\pi\mu_0IN/L$. Again we assume that the \tmu functions are constant
throughout the size of the experimental  device. In terms of fundamental
atomic parameters, $L$ is proportional to an integer times the Bohr
radius (the interatomic spacing), which is known to rescale as
$a_0\to a_0\epsilon_0c_0^{2}/\sqrt{T_0}$. If we now
write $I=\int \vec J\cdot d\vec S$, where $J$ can be expressed
in terms of a density charge $\rho$ in motion ($v$) through a
volume $V$, and then relate the Bohr 
radius to  each spatial dimension along with
the limiting particle velocity $c_0$ to the velocity distribution $v$,
we can show $I\to I\sqrt{T_0}/\epsilon_0 c_0$, and so
$B\to B\mu_0 T_0/\epsilon_0^2c_0^3$. Along with (\ref{rescal}), this gives 
the position dependence of (\ref{anor1}) to be
\be\label{anor2}
\om_a=\om_a^0\sqrt{T_0}\frac{\mu_0c_0}{(\epsilon_0 c_0)^3}(1+\xi\frac{7}{6})
\ee
with $\om_a^0=eB\al/2m\pi$  (recall $\xi=1-1/\mu_0\epsilon_0 c_0^2$).

Note that the \tmu functions are evaluated at some representative point
of the system, which we have chosen to be the origin $\vec X=0$. 
In order to determine how
$\om_a$ changes as the position of the system varies, we expand
the \tmu functions in (\ref{anor2}) 
\be\label{tu}
T(U)=T_0+T_0'\vec g_0\cdot\vec X+O(\vec g_0\cdot\vec X)^2
\ee
where $\vec g_0=\vec\gr U|_{\vec X=0}$, $T_0=T|_{\vec X=0}$, and
$T_0'=dT/dU|_{\vec X=0}$. It is useful to redefine the gravitational
potential $U$ by
\be\label{u}
U\to-\half\frac{T_0'}{H_0}\vec g_0\cdot\vec X
\ee
whose gradient yields the test-body acceleration $\vec g$. This finally
yields
\be\label{anor3}
\om_a=\om_a^0\Big[1-U+(\frac{11}{6}\Gamma_0-\frac{13}{6}\Lambda_0)U\Big]
\ee
where we have rescaled again according to (\ref{rescal}), and omitted terms
proportional to $\xi$, since the main
position dependence parameterization is given in terms of:
\be\label{gamlam}
\Gamma_0=\frac{2T_0}{T_0'}(\frac{\epsilon_0'}{\epsilon_0}+
\frac{T_0'}{2T_0}-\frac{H_0'}{2H_0}),
\qquad
\Lambda_0=\frac{2T_0}{T_0'}(\frac{\mu_0'}{\mu_0}+
\frac{T_0'}{2T_0}-\frac{H_0'}{2H_0})
\ee

By comparing eq. (\ref{anor3}) with (\ref{para1}), we can 
identify 
\be
\Xi^{g-2}=\frac{11}{6}\Gamma_0-\frac{13}{6}\Lambda_0
\ee
as the LPI-violating parameter. Note that this
depends on the anomalous frequency related to 
the longitudinal polarization of the beam.  It is also
species-dependent, with the value of $\Gamma_0$ and $\Lambda_0$
for the electron differing from that of the muon. A search
for possible position dependence of anomalous spin precession
frequencies provides another qualitatively new test of
LPI sensitive to radiative corrections.

Actually the most precise $g-2$ experiments for electron 
measure the ratio $a=\om_a/\om_c$ at non relativistic
electron energies ($\beta\sim 10^{-5}$), and so $\om_c\simeq
eB/m$. This is interesting because by following the
former parameterization we can write:
\be\label{ciclo}
\om_c=\om_c^0\Big[1-U+(2\Gamma_0-\Lambda_0)U\Big]
\ee
or by taking the ratio of (\ref{anor3}) to (\ref{ciclo}):
\be
a=a^0(1+U\Xi^a),\qquad \Xi^a=\frac{1}{6}\Gamma_0+\frac{7}{6}\Lambda_0
\ee
and then by identifying $a$ with the most precise experimental value \cite{vandyck}
and $a^0$ with the theoretical one\cite{Kino}, we can constrain
through the resulting theoretical/experimental 
errors $|U\Xi^a|<3\times 10^{-8}$. This result is sensitive to the
absolute value of the total local gravitational potential
\cite{Hughes}, whose
magnitude has recently been estimated to be as large as
$3\times 10^{-5}$ due to the local supercluster \cite{Kenyon}.
Hence measurements of this type can provide us with empirical
information sensitive to radiative corrections that
constrains the allowed regions of $(\Gamma_0,\Lambda_0)$ parameter
space, giving in this case: 
\be
|\frac{1}{6}\Gamma_0+\frac{7}{6}\Lambda_0|<10^{-3}
\ee
For muons the analogous constraint is $|U\Xi^a_\mu|< 10^{-5}$, and so 
nothing conclusive is obtained.

We pause to compare this result to an analogous result obtained
for hyperfine transitions (maser clocks). In this
case the baryonic and leptonic
gravitational parameters appear  simultaneously. The 
atomic hyperfine splitting comes from the interaction between the
magnetic moments of the electron and proton (nucleus). The proton metric
appears only in the latter, and so it does not affect the principal
and fine structure atomic energy levels.  Non-metric effects imply
a shift in the hyperfine energy $E_{hf}$ which is \cite{will1}
\be\label{hf1}
\Delta E_{hf}={\cal E}_{hf}(1-U_B)+{\cal E}_{hf} U_B\Xi^{hf}
\ee
with
\be
\Xi^{hf}=3\Gamma_B-\Lambda_B+\Delta
\ee
where $U_B$, $\Gamma_B$ and $\Lambda_B$ are the baryonic analogues of
the parameters appearing in (\ref{gamlam}).
In (\ref{hf1}) we rescaled the atomic parameters to absorb the
\tmu functions and chose units such that $c_B=1$. The quantity $\Delta$ is
given by 
\be
\Delta=2\frac{T_B}{T_B'}\left[2(\frac{H_B'}{H_B}-\frac{H_0'}{H_0})
-\frac{T_B'}{T_B}+\frac{T_0'}{T_0}\right]
\ee
and would vanish under the assumption that the leptonic and baryonic
\tmu parameters were the same. The gravity probe A experiment 
\cite{redshift}, employing hydrogen-maser
clocks, was able to constrain the corresponding LPI violating
parameter related to hyperfine transitions, obtaining 
\begin{equation}
|\Xi^{Hf}|=|3\Gamma_B-\Lambda_B +\Delta|<2\;\times\;10^{-4}
\label{hyplpi}
\end{equation}
for the most stringent bound to date on $\Xi^{Hf}$.

We note that a similar analysis could be carried out for the 
energy shift defined in (\ref{eshift}),
which can be used as a frequency test to look for position 
or frame dependence. This can
be done by following the same procedure as for atomic energy
 shifts, where the anomalous
passive and inertial gravitational tensor are introduced in 
order to relate non-metric effects
to redshift and time dilation parameters. Since radiative 
corrections are irrelevant in that
energy shift, we omit that procedure here.

\section{Concluding Remarks}

Refined measurements of anomalous magnetic moments can provide
an interesting new arena for investigating the validity of the EEP
in physical systems where radiative corrections are important.
We have considered this possibility explicitly for the class of
non-metric theories described by the \tmu formalism.
The non-universal character of the gravitational couplings in such
theories  affects
the one loop corrections to the scattering amplitude of a free
fermion in an external electromagnetic field
in a rather complicated way, giving rise to several novel effects.

An evaluation of the one-loop diagrams reveals that 
the leg corrections, which in the metric case give
no contribution to the total amplitude 
after a proper renormalization of mass and spinor field,
provide contributions which 
cannot be removed after renormalization. Moreover they are essential
in ensuring the gauge invariance of the scattering
amplitude, which is not fulfilled by the vertex correction alone. 
The consistency of the
calculation is verified explicitly through the Ward identity, which 
furnishes a
cross-check between the fermion self energy and the vertex correction.
The non metric corrections to the scattering amplitude also have
an infrared divergence, which could be 
understood in terms of inelastic soft
photon radiation, as in the metric case. 
This does not affect the term associated with
the anomalous magnetic moment. 

The presence of preferred frame effects induces a new type of coupling
between the magnetic field and the spin as described by (\ref{inter}).
This interaction stems purely from radiative corrections,
and generalizes the gyromagnetic ratio of a fermion to a tensorial 
coupling described by $\Gamma_{ij}$. 
We emphasize that qualitatively new information on the validity of
the EEP will be obtained by  setting new empirical bounds on this
coupling, as it is associated with purely
{\it leptonic} matter. Comparatively little is known about such
empirical limits on EEP-violation relative to the baryonic sector
\cite{Hughes}, for which previous experiments have set the limit
\cite{PLC} $|\xi_B| \equiv |1 -c_B^2|\,<\,6\,\times \,10^{-21}$
where $c_B$ is the ratio of the limiting speed of baryonic matter
to the speed of  light. We can therefore safely neglect
any putative effects of $\xi_B$ in our analysis. 

Consequently, discussion of a $g-2$ contribution to the magnetic moment 
no longer makes sense, and we instead refer to the anomalous frequency as
the main connection with experiment. Note that this frequency, defined as
the relative electron spin precession with respect to its velocity, 
comes from radiative corrections and it becomes proportional to $g-2$ in
the metric case. This frequency shows an explicit dependence on both
the preferred frame velocity and its relative direction with respect to the
external magnetic field. There is also a dependence 
on the electron velocity,
which makes the other contributions negligible at relativistic electron
energies. Two $g-2$ experiments on the electron (one at
relativistic energies and the other almost at rest) may then be used
to limit the preferred
frame parameter to be no larger than $10^{-5}$, analogous to the work of 
Newman {\it et al.}.  Constraining any possible EEP violation to
be no larger than the present discrepancy between theory and experiment 
we found the most stringent bounds for $\xi$ yet obtained
for leptonic matter, as given in (\ref{belectron}) and (\ref{bmuon}).

We expect that new experiments which probe
the anisotropic character (or angular dependence) of the frequency 
could be used to impose stronger limits in different physical regimes.
For example, as the Earth rotates, the spatial orientation of the
magnetic field changes -- this should in turn diminish the
experimental errors involved in the comparison between  two 
energetically different $g-2$ experiments.

The relativistic generalization of the spin polarization equation 
(\ref{spin5}), followed the same procedure as for the metric case, 
where non-metric effects where included
in the interaction only (eq. (\ref{inter})). This yields an 
equation of motion for the
spin (as seen from the rest frame) which is qualitatively different from 
that expected from its classical counterpart,
where the angular momentum rate is related to the torque 
applied on the system.
This approach for dealing with violations of Lorentz invariance is 
dynamical; from a kinematical
viewpoint we assume that standard Lorentz transformations 
relate coordinates and fields from one system to another.

Perhaps the most
remarkable feature of the non-metric effects 
is that of the oscillations of the component of spin
polarization parallel to the magnetic field. Since this component 
remains constant in the metric case, an experiment which searches for
such oscillations is a new null test of the equivalence principle that
is uniquely sensitive to radiative corrections in the leptonic sector.
Hence an empirical investigation
of its behavior will provide qualitatively new information about the
validity of EEP, and could
constrain even further the limits on the preferred
frame parameters.

Finally, we analyzed the behavior of 
the anomalous frequency in the context of
redshift experiments, which can put constraints on
the LPI-violating parameters ($\Gamma_0$, $\Lambda_0$) once the
corresponding experiments are carried out. This region of parameter
space is qualitatively different from that probed by either Lamb-shift
or hyperfine effects. In the electron sector a bound on the magnitude
of $U\Xi^a$ can be obtained by demanding that it be no larger than
the error bounds in the discrepancy between the experimental and
theoretical values of the ratio $a=\om_a/\om_c$. Assuming the local
potential to be as large as that estimated from the local supercluster,
we obtain a bound on $|\Xi^a|$ that is comparable the limit on
an analogous quantity in the baryonic sector obtained from redshift
experiments \cite{redshift}.  However this latter experiment is 
proportional to changes in the local potential, which are $\sim 10^{-10}$.
More direct limits on $|\Xi^a|$ must be set by performing a similar
sort of redshift experiment on anomalous magnetic moments 
\cite{gasp}. The logistics and higher precision demanded by such
an experiment will be a major challenge to undertake.

\section*{Acknowledgments}

This work was supported in part by the Natural Sciences and Engineering
Research Council of Canada.
 
\setcounter{equation}{0}
\setcounter{section}{-1}
\begin{center}
\section{ Appendices}
\end{center}

\renewcommand{\thesection}{\Alph{section}}
\renewcommand{\theequation}{\Alph{section}\arabic{equation}}
\section{ Loop integrations}

We show the main steps leading to Eqs. (\ref{vertex-f}), (\ref{self-f}), and (\ref{pola-f}).
Details are given throughout the computation by considering only the first term of the photon propagator 
(\ref{fp}),  that is
\be\label{fp1}
G_{\mu\nu}^\xi=\frac{\beta_\mu\beta_\nu}{k^2}+\cdots
\ee
with the remaining term in (\ref{fp}) contributing in a similar manner.

We solve for the fermion self energy by replacing (\ref{fp1}) in (\ref{self}), and using
(\ref{regu}) along with the Feynman parameters
$$ \frac{1}{a^2b}=2\int_0^1\frac{dz\, z}{[az+b(1-z)]^3}.$$ After 
integrating we obtain
\be
\Sigma(p)=\frac{\al}{4\pi}\int\nn\beta(\nn p z+m)\nn\beta\ln\left(\frac{
m^2(1-z)^2+z\Lambda^2-\Delta z(1-z)}{m^2(1-z)^2-\Delta z(1-z)+z\mu^2}\right)+\cdots
\ee
with $\Delta=p^2-m^2$. We consider $\Delta/m^2<<1$, and expand the above
to obtain after some manipulation 
\ba
&&\Sigma(p)=\frac{\al}{4\pi}\Big\{\beta\cdot p\nn\beta\left(\frac{5}{2}+\ln(\frac{\Lambda}{m})^2
\right)-(\nn p-m)\frac{\beta^2}{2}\left(\ln(\frac{\Lambda}{m})^2+\half\right)\\\nonumber
&&-m\frac{\beta^2}{2}\left(\frac{1}{2}-\ln(\frac{\Lambda}{m})^2\right)
+2\frac{(\beta\cdot p)^2}{m^2}(\nn p-m)\left(\ln(\frac{m}{\mu})^2-3\right)\Big\}
+O\Big((\nn p-m)^2\Big)+\cdots
\ea
where we have kept the leading terms as $\mu\to 0$ and $\Lambda\to\infty$, and
$O\Big((\nn p-m)^2\Big)$ stands for the terms  satisfying
$$ S_F(p)O\Big((\nn p-m)^2\Big)u(\vec p)=0.$$

We renormalize $\Sigma(p)$ by subtracting 
\be
\delta \Sigma=\delta m+\delta\xi_0\nn\beta\beta\cdot p\ee
where the counterterms respectively
account for mass and \tmu-parameter renormalization.

Choosing the counterterms so that
$$\ov u(\vec p)\Sigma(p)u(\vec p)=0,$$
and so
\ba
\delta m&=&-\frac{\al}{4\pi}m\frac{\beta^2}{2}\left(\frac{1}{2}-\ln(\frac{\Lambda}{m})^2\right)\\
\label{xi0}\delta\xi_0&=&\frac{\al}{4\pi}\left(\frac{5}{2}+\ln(\frac{\Lambda}{m})^2
\right)
\ea
the regularized result is then
\ba
\Sigma(p)&=&\api(\nn p-m)\left\{\frac{(\beta\cdot p)^2}{m^2}\left[\half \ln(\frac{m}{\mu})^2
-\frac{3}{2}\right]-\beta^2\left[\frac{1}{8}\ln(\frac{\Lambda}{m})^2+\frac{1}{16}\right]
\right\}\nonumber\\ 
&+&O\Big((\nn p-m)^2\Big)\quad .
\ea
Note that the remaining ultraviolet divergence related to this term could be
removed after charge renormalization. We find it convenient to leave it
in order to cross-check the calculation, since a similar term from the vertex
part should cancel it, thereby
removing the divergence from  the resulting scattering amplitude.

The evaluation of the vertex function follows 
a similar procedure, giving the result
\ba\label{vertex-a}
\Gamma^\mu&=&\api\int \frac{dx}{p_x^2}\Big\{\gamma^\mu\Big[\half\beta\cdot p\,\beta\cdot p'
(\ln\frac{p_x^2}{\mu^2}-2)+\frac{\beta^2}{8}p_x^2(\frac{3}{2}-\ln\frac{\Lambda^2}{p_x^2})\Big]\nonumber\\
&-&x\frac{\nn p'}{2}\Big(\beta\cdot p\gamma^\mu\nn\beta+\beta\cdot p'\nn\beta\gamma^\mu\Big)
-\Big(\beta\cdot p\gamma^\mu\nn\beta+\beta\cdot p'\nn\beta\gamma^\mu\Big)(1-x)\frac{\nn p}{2}\\
&+&\nn\beta\Big(\half p_x\cdot\beta p_x^\mu-\frac{1}{4} \beta^\mu p_x^2(\frac{3}{2}
 -\ln\frac{\Lambda^2}{p_x^2})
-(1-x)\beta\cdot p\,  p^\mu-x p'\cdot\beta\,  p'\,^\mu\Big)\nonumber\\
&+&\nn p_x\Big(\beta^\mu(\beta\cdot p +\beta\cdot p')-\frac{\beta^2}{4}p_x^\mu\Big)
\Big\}+\cdots\nonumber
\ea
with $p_x=xp'+(1-x)p$.
Since (\ref{vertex-a}) is acting on a free spinor, we can use
$$p_x^2=m^2-x(1-x)q^2,$$
with $q=p'-p$, and so expand
$$\frac{m^2}{p_x^2}=1+x(1-x)\frac{q^2}{m^2}+O(q^4),$$
which after some algebra reduces (\ref{vertex-a}) to
\ba
 \Gamma^\mu(p',p)&=&\api\Big\{\gamma^\mu\left[-\beta^2\left(\frac{1}{16}+\frac{1}{8}
\ln(\frac{\Lambda}{m})^2\right)+\frac{(\beta\cdot p)^2}{m^2}
\left(\half\ln(\frac{m}{\mu})^2-\frac{3}{2}\right)\right.\nonumber\\
&+&\left.\frac{q^2}{m^2}\left(\frac{1}{12}
\ln(\frac{m}{\mu})^2-\frac{1}{3}-\frac{\beta^2 }{16}\right)+\frac{\beta\cdot p}{m}\frac{\beta\cdot q}{m}
\left(\half\ln(\frac{m}{\mu})^2-\frac{5}{4}\right)\right]\nonumber\\
&+&\frac{\nn q}{m}\gamma^\mu\nn\beta\left(\frac{1}{4}\frac{\beta\cdot p}{m}+\frac{1}{8}\frac{\beta\cdot q}{m}
\right)+\frac{5}{24}\frac{q^2}{m^2}\nn\beta\beta^\mu-\half\frac{\nn q}{m}\nn\beta\beta^\mu\\
&-&\half\frac{\beta\cdot q}{m}\nn\beta\gamma^\mu+\frac{\beta\cdot q}{m}\beta^\mu+
\frac{\beta^2}{8}\frac{\nn q}{m}\gamma^\mu-\left(\frac{1}{4}\frac{\beta\cdot p}{m}+\frac{1}{3}\frac{\beta\cdot q}{m}
\right)\frac{q^\mu}{m}\nn\beta\nonumber\\
&-&\frac{\beta^2}{8}\frac{q^\mu}{m}\Big\}+O(q^3)\nonumber
\ea
where  the vertex function has been renormalized by subtracting a term like
\be
\delta\Gamma^\mu=\delta\xi_0\nn\beta\beta^\mu
\ee
with $\delta\xi_0$ is given by (\ref{xi0}). We recall that gauge invariance
forces this coefficient to be equal to the one participating in the renormalization
of the fermion self energy.

\setcounter{equation}{0}
\section{Adiabatic hypothesis}

In order to describe how self energy effects convert the incident electron from a bare
particle to a physical one, it is convenient to introduce a damping function, $g(t)$,
which adiabatically switches off the coupling between fields, such that the interaction
lagrangian is replaced by
\be
{\cal L}_I=eg(t)\sp(x)\nn A(x)\psi(x)
\ee
It is assumed that the time $T$ over which $g(t)$ varies is very long compared to the
duration of the scattering process.  In momentum space
\be
g(t)=\int G(\om_0)\exp(i\om\cdot x)d\om_0
\ee
with $\om\equiv(\om_0,0)$, and $g(0)=1$. It is supposed that $G(\om_0)$ is almost a
delta function, being large for $ \om_0$ in a range of about $T^{-1}$

In the presence of an external field $A_\mu$, eq. (\ref{leg}) will now read
\be
L\cdot A\rightarrow\int G(\om_0)G(\om_0')d\om_0d\om_0'
\nn A(p'-p-\om-\om')iS_F(p-\om-\om')i\Sigma(p-\om)+\cdots
\ee
where $\cdots$ represents the equivalent second term from (\ref{leg}).

As $T\to\infty$, and $\om_0$, $\om_0'\to 0$, the fermion propagator reduces to
\be
S_F=\frac{1}{\nn p-\nn\om-\nn\om '-m}\simeq-\frac{\nn p+m}{2p_0(\om_0+\om_0')}
\ee
where we used $p^2=m^2$. This implies that  we can expand $\Sigma$ up to
order $\om$ only, since higher terms vanish after taking the previous limit. 
Here we employ the relation
\be
\frac{1}{A-B}=\frac{1}{A}+\frac{1}{A}B\frac{1}{A}+\cdots
\ee
to expand
\be\label{sig1}
\Sigma(p-\om)\simeq\Sigma(p)-\frac{\pd\Sigma(p)}{\pd p^\mu}\om^\mu
\quad .
\ee
After renormalization, $\Sigma$ takes the form
\be\label{sig2}
\Sigma(p)=(\nn p-m)(A+B(\beta\cdot p)^2)+C(\beta\cdot p)\Big\{
\nn\beta(\nn p-m)+(\nn p-m)\nn\beta\Big\}+\cdots
\ee
where the constants $A$, $B$, and $C$ can be obtained from
eq. (\ref{self-f}).

Let us introduce $\ov\om\equiv\om+\om'$, and symmetrize $\om$ by
$\half(\om+\om')$ in (\ref{sig1}), to write
\be
S_F(p-\om-\om')\Sigma(p-\om)=\frac{1}{\nn p-\nn\ov\om-m}
(\Sigma(p)-\half\frac{\pd\Sigma(p)}{\pd p^\mu}\ov\om^\mu)
\ee
which after using (\ref{sig2})  can be written as
\be\label{sig3}
\half(A+B(\beta\cdot p)^2)+C\beta\cdot p\nn\beta
\ee
where we have used that $\Sigma(p-\om)$ is acting on a free
spinor, and therefore terms of the form $(\nn p-m)u(p)$ vanish.
Now, the final evaluation of (\ref{leg}) follows directly from (\ref{sig3}).


\begin{thebibliography}{99}

\bibitem{PLC} J. D. Prestage et al., Phys. Rev. Lett. {\bf 54},  2387 (1985);
 S. K. Lamoreaux et al.,  {\it ibid.} {\bf 57}, 3125 (1986);
T. E. Chupp et al., {\it ibid.} {\bf 63}, 1541 (1989).

\bibitem{will1} C. M. Will, {\it Theory and Experiment in Gravitational Physics},
2nd edition (Cambridge University Press, Cambridge, 1992).

\bibitem{HW} M. P. Haugan and C. M. Will, Phys. Rev. Lett. {\bf 37}, 1 (1976);
Phys. Rev D. {\bf 15}, 2711 (1977)

\bibitem{Hughes} R.J. Hughes, Contemporary Physics {\bf 34} 177 (1993).

\bibitem{Jody}M. Gabriel, M. Haugan, R.B. Mann and J. Palmer,
Phys. Rev. Lett. {\bf 67} (1991) 2123. 

\bibitem{anti} M.H. Holzscheiter, T. Goldman and M.M. Nieto 
Los Alamos preprint LA-UR-95-2776, (Sept. 1995),
{\tt hep-ph/9509336}.

\bibitem{utpal}R.B. Mann and U. Sarkar,
Phys. Rev. Lett. {\bf 76} (1996) 865. 

\bibitem{Kenyon}I.R. Kenyon, Phys. Lett. {\bf B237} 274 (1990).

\bibitem{catlamb}C. Alvarez and R.B. Mann, ``Test of the Equivalence
Principle by Lamb Shift Energies'' WATPHYS-TH95/02, gr-qc/9507040.

\bibitem{NFRS} D. Newman, G. W. Ford, A. Rich, E. Sweetman, Phys. Rev. Lett. {\bf 40},
1355 (1978) 

\bibitem{tmu} A. P. Lightman and D. L. Lee, Phys. Rev. D {\bf 8}, 364 (1973).

\bibitem{gabriel} M. D. Gabriel and M. P. Haugan, Phys. Rev. D {\bf 41}, 2943 (1990).

\bibitem{M&S} F. Mandl, G. Shaw, {\it Quantum Field Theory} (Wiley, Toronto, 1984).

\bibitem{haugan} M. P. Haugan, Ann. Phys. (N.Y.) {\bf 118}, 156 (1979).

\bibitem{Kino}  T. Kinoshita and D. R. Yennie, {\it High Precision Test of Quantum
Electrodynamics-An Overview}, ed T. Kinoshita (World Scientific, Singapure,1990), p. 1.

\bibitem {Bailey}J. Bailey {\it et.al.}, Phys. Lett. {\bf B68}, 191 (1977). 

\bibitem{Bailey2}J. Bailey {\it et.al.}, Nuovo Cimento {\bf A9}, 369 (1972).

\bibitem{vandyck} R. Van Dyck, P. Schwinberg, and H. Dehmelt, Phys. Rev. Lett. {\bf 59},
26 (1987) 

\bibitem{redshift} R. F. C. Vessot and M.W. Levine, Gen. Relativ. Gravit. {\bf 10},181 (1979).

\bibitem{gasp}M. Gasperini, Mod. Phys. Lett. {\bf A4}, 1681 (1989).

\end{thebibliography}
\end{document}